\documentclass{emulateapj}
\newcommand{\co}{$^{12}$CO~}
\newcommand{\tco}{$^{13}$CO~}
\newcommand{\kms}{km s$^{-1}$~}

\newcommand{\kmsc}{km s$^{-1}$}

\newcommand{\ad}{${\arcdeg}$}

\shorttitle{Triggered Star Formation}
\shortauthors{Kang et al.}

\begin{document}
\title{Triggered Star Formation in a Double Shell near W51A}
\slugcomment{Accepted for publication in ApJ}
\author{\sc Miju Kang\altaffilmark{1,2,3}, 
            John H. Bieging\altaffilmark{3}, 
            Craig A. Kulesa\altaffilmark{3},
            and Youngung Lee\altaffilmark{1}}
\altaffiltext{1}{International Center for Astrophysics,
                 Korea Astronomy and Space Science Institute,
                 Hwaam 61-1, Yuseong, Daejeon 305-348, South Korea;
                 mjkang@kasi.re.kr.}
\altaffiltext{2}{Department of Astronomy and Space Science,
                 Chungnam National University, Daejeon 305-764, South Korea}
\altaffiltext{3}{Steward Observatory,
                 University of Arizona,
                 933 North Cherry Avenue, Tucson, AZ 85721}

\begin{abstract} 

We present Heinrich Hertz Telescope CO observations of the shell
structure near the active star-forming complex W51A to investigate the
process of star formation triggered by the expansion of an \ion{H}{2}
region. The CO observations confirm that dense molecular material has
been collected along the shell detected in {\it Spitzer} IRAC images.
The CO distribution shows that the shell is blown out toward a lower
density region to the northwest. Total hydrogen column density around
the shell is high enough to form new stars. We find two CO condensations
with the same central velocity of 59 \kms to the east and north along
the edge of the IRAC shell. We identify two YSOs in early evolutionary
stages (Stage 0/I) within the densest molecular condensation. From the CO
kinematics, we find that the \ion{H}{2} region is currently expanding with
a velocity of 3.4 \kmsc, implying that the shell's expansion age is $\sim$
1 Myr. This timescale is in good agreement with numerical simulations
of the expansion of the \ion{H}{2} region (\citeauthor{Hosokawa06}).
We conclude that the star formation on the border of the shell is
triggered by the expansion of the \ion{H}{2} region.

\end{abstract}


\section{Introduction}

Feedback from massive stars has significant impact on the surrounding
interstellar medium (ISM), and may even trigger star formation in the
vicinity of their expanding \ion{H}{2} regions \citep{Elmegreen77}.
An expanding \ion{H}{2} region sweeps up an ambient interstellar medium
and accumulates the material between the ionization front and the
shock front. The material collected on the border of the \ion{H}{2}
region will form a shell structure and would appear as a ring on the
plane of the sky. Second-generation stars may form when the compressed
layer becomes gravitationally unstable or the expanding shell provides
enough external pressure to initiate collapse of a pre-existing molecular
clump. There are relatively few good examples of this process published to
date \cite[e.g.,][]{Deharveng03,Zavagno06,Watson08,Pomares09}. Therefore,
identifying other cases of shell structures around OB stars, together
with detailed study of the associated interstellar matter, will provide
valuable understanding of the triggered star formation process.

High spatial resolution of the Spitzer Space Telescope makes it
possible to routinely detect these shell structures at mid-infrared
wavelengths. \cite{Churchwell06} found 322 shell structures
from the Galactic Legacy Infrared Mid-Plane Survey Extraordinaire
\cite[GLIMPSE;][]{Benjamin03}. They reported that many of the infrared
shells coincide with known \ion{H}{2} regions produced by O and early-B
type stars. About 7\% of the shells in their catalog reveal a multiple
morphology. The shell structure we present in this paper is a multiple
bubble consisting of N102 and N103 in their catalog.

In this paper, we use infrared and CO observations to study the
double-shell structures and their interactions with the surrounding ISM.
Infrared observations from Spitzer show the morphology of the shell and
reveal the YSOs possibly formed by triggering in the surrounding ISM.
Direct observation of molecular gas around the shell is necessary (1)
to confirm that ambient molecular material is indeed associated with the
expanding \ion{H}{2} region, (2) to find any clumpy material that could
condense to form new stars, and (3) to learn the physical properties
(e.g., density of accumulated molecules) of the ambient ISM into which
the shell is expanding.  These observations will enable us to determine
whether or not it is possible to form stars by the triggering effect of
the \ion{H}{2} region.

This paper is organized as follows. We introduce our CO observations
and the data sets used in this paper in \S\,\ref{sec:observation}. In
\S\,\ref{sec:morphology}, we describe the morphology of the N102 and N103
region. We identify the ionizing star in \S\,\ref{sec:ionizing_star} and
any young stellar objects (YSOs) around the shell in \S\,\ref{sec:YSO}.
In \S\,\ref{sec:CO} we present the distribution and kinematics
(\S\,\ref{sec:distribution_kinematics}) and physical conditions
(\S\,\ref{sec:physical_condition}) of the molecular cloud associated
with the shell. We discuss the triggered star formation on the border
of the \ion{H}{2} region in \S\,\ref{sec:discussion}. We summarize our
results in \S\,\ref{sec:conclusion}.\\

\begin{figure}[htb!]
\epsscale{1.2}
\plotone{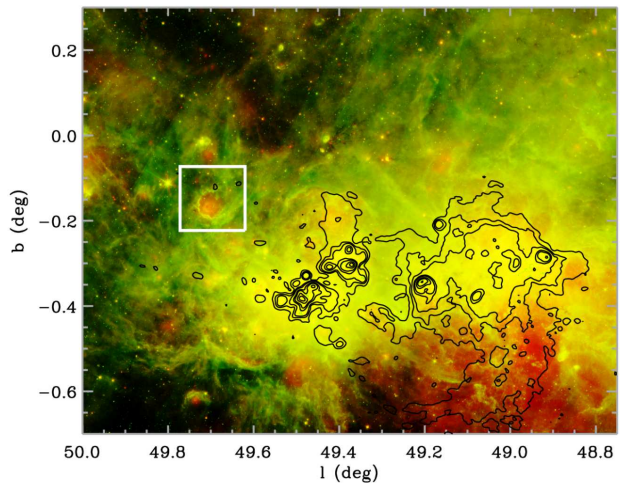}

\caption{ A color image of the W51 complex composed of MIPS 24 $\micron$
(red) and IRAC 8.0 $\micron$ (green). The contour map shows 21 cm radio
continuum emission. It was made by combining VLA data and Effelsberg 100-m
telescope data \citep{Koo97}. The contour levels are $-$0.015, 0.015,
0.05, 0.10, 0.3, 0.5, 1.0, 1.5, and 2.4 Jy beam$^{-1}$. The white box
shows the outline of the shell area (Figure \ref{shell_rgb_irac_mips}).}

\label{w51_shell}
\end{figure}

\section{Observations and Data Analysis}
\label{sec:observation}

We have carried out \co and \tco (J=2-1) line observations of the W51
\ion{H}{2} region complex with the 10 meter Heinrich Hertz Telescope (HHT)
on Mt. Graham, Arizona. The whole map covers a 1.25$\arcdeg$ $\times$
1.00$\arcdeg$ region centered at ($l,b$) = ($49.375\arcdeg$, $-
0.2\arcdeg$). In this paper, we extract a 0.152$\arcdeg$ $\times$
0.150$\arcdeg$ region centered  at ($l,b$) =(49.696\arcdeg,
$-$0.148\arcdeg) which includes the shell structure. The \co and
\tco (J=2-1) line observations were mapped with on-the-fly (OTF)
scanning in longitude at 10$\arcsec$ per second with row spacing of
10$\arcsec$ in latitude. We used the 1.3mm ALMA band 6 dual polarization
sideband-separating receiver with a 4-6 GHz IF band. The receiver was
tuned with the \co (J=2-1) line at 230.538 GHz in the upper sideband and
\tco (J=2-1) line at 220.399 GHz in the lower sideband. The spectrometers,
one for each of the two polarizations and the two sidebands, were filter
banks with 256 channels of 1 MHz width and separation. Beam efficiency
was 0.85, which we adopt for all the data.

Data for each CO isotopomer were processed with the CLASS reduction
package (from the University of Grenoble Astrophysics Group), by
removing a linear baseline and convolving the data to a square grid
with 10$\arcsec$ grid spacing. The intensity scales for the two
polarizations were determined from observations of W51D made just
before the OTF maps.  System temperatures were calibrated by the
standard ambient temperature load method \citep{Kutner81} after every
other row of the map grid. Further analysis was done with the Miriad
software package \citep{Sault95}. Although the data observed early
in the program with a different receiver were noisier than with the
ALMA receiver, data quality was improved after combining all of the
data using ($1/T_{sys}$)$^{2}$ weighting. To further reduce the noise
of the images without significantly sacrificing angular resolution, we
convolved the maps with a 16$\arcsec$ (FWHM) Gaussian. This increases the
effective resolution from 32$\arcsec$ to 36$\arcsec$, but smooths over
the 10$\arcsec$ OTF raster pattern, improving the final image quality. The
two polarizations were averaged for each sideband, yielding images with
rms noise per pixel and per velocity channel of 0.16 and 0.07 K-T$_A^*$
for the \co and \tco transition respectively. The average rms noise
level of the \co map is larger than \tco because several sub-fields of
\co map were observed just one time with the noisier receiver.

\tco (J=1-0) data covering the same region as our mapping
area were extracted from the Galactic Ring Survey (GRS)
\footnote{http://www.bu.edu/galacticring/} with spectral resolution of 0.2
\kmsc, angular resolution of 46$\arcsec$ and sampling of 22$\arcsec$. The
rms noise level of the GRS data when smoothed to 1.3 \kms velocity
resolution is 0.08 K-T$_A^*$. The intensities on an antenna temperature
scale were divided by the main beam efficiency of 0.48 to convert to
main-beam brightness temperature.

The Galactic Legacy Infrared Midplane Survey Extradodinaire (GLIMPSE)
I survey observed the Galactic plane ($65\arcdeg < |l| < 10\arcdeg$
for $ |b| < 1\arcdeg$) with the four mid-IR bands (3.6, 4.5, 5.8, and
8.0 $\micron$) of the Infrared Array Camera \cite[IRAC;][]{Fazio04} on
the Spitzer Space Telescope. The Midinfrared Imaging
Photometer for Spitzer Galactic plane Survey \cite[MIPSCAL;][]{Carey05}
is a legacy program covering the inner Galactic plane, $65\arcdeg <
|l| < 10\arcdeg$ for $ |b| < 1\arcdeg$, at 24 and 70 \micron\ with the
Multiband Imaging Photometer for Spitzer \cite[MIPS;][]{Rieke04} on the
Spitzer Space Telescope. We analyzed mosaicked images of four IRAC bands
and MIPS 24 $\micron$. The resolutions of IRAC and MIPS 24 $\micron$
are 1\farcs2 and 2\farcs4, respectively.

We use the GLIMPSE I Catalog which consists of point sources that
are detected at least twice in one band with a S/N $>$ 5. The GLIMPSE
I Catalog also includes $JHK_{s}$ flux densities from the 2MASS point
source catalog \citep{Skrutskie06}. MIPS 24 $\micron$ fluxes were
obtained from running the GLIMPSE pipeline on the MIPSGAL BCD mosaics
(Meade, private communication). We extract 24 $\micron$ point-sources
with $S/N > 7$, then bandmerge the 24 $\micron$ sources with the GLIMPSE
Catalog sources using a 2\farcs0 correlation radius. The final source
list consists of all 8 bands combined: 2MASS $JHK_{s}$, 
IRAC (3.6, 4.5, 5.8, and 8.0\micron), and MIPS 24 $\micron$.

\begin{figure}
\epsscale{1.20}
\plotone{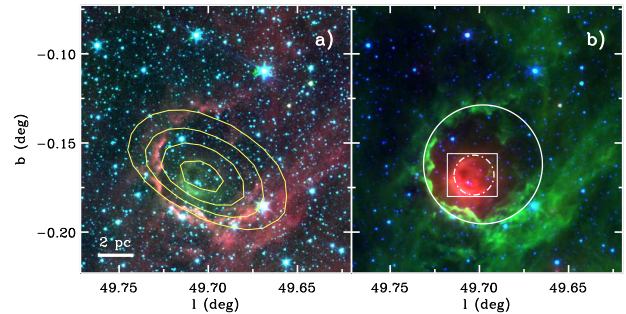}

\caption{({\it a}) Composite image of IRAC 8.0 $\micron$ (red), IRAC
4.5 $\micron$ (green), and IRAC 3.6 $\micron$ (blue). Contours show
Westerbork Synthesis Radio Telescope(WSRT) 327 MHz radio continuum
\citep{Taylor96}. Contour levels are 0.07, 0.12, 0.16, and 0.21 Jy
beam$^{-1}$. Scale bar on the left bottom represents a size at the
distance of 5.7 kpc. ({\it b}) Composite image of MIPS 24 $\micron$ (red),
IRAC 8.0 $\micron$ (green), and IRAC 4.5 $\micron$ (blue).  Outer solid
circle represents a shell size of 2$\arcmin$ radius centered on ($l,
b$) = (49.698$\arcdeg$, -0.162$\arcdeg$) of N102. Inner dash-dotted
circle represents N103 with 0.67$\arcmin$ radius centered on ($l, b$) =
(49.703$\arcdeg$, -0.168$\arcdeg$) \citep{Churchwell06}. The white box
is the outline enlarged in Figure \ref{shell_rgb_irac_cent}.\vspace{0.5cm}}

\label{shell_rgb_irac_mips}
\end{figure}

\section{Properties of the Region} 
\subsection{Spitzer Images and Radio Continuum}
\label{sec:morphology}

Figure \ref{w51_shell} shows a two-color composite image of Spitzer IRAC
8 $\micron$ (green) and MIPS 24 $\micron$ (red) data from the GLIMPSE
and MIPSGAL Legacy data bases, for a 1\arcdeg $\times$ 1.25 \arcdeg
area including the W51 star-forming region. The contours outline the
radio continuum emission from the ionized gas and supernova remnant
\citep{Koo97}. The double shell structure studied in this paper is located
in the white boxed area of Figure \ref{w51_shell}.

\begin{figure}
\epsscale{1.1}
\plotone{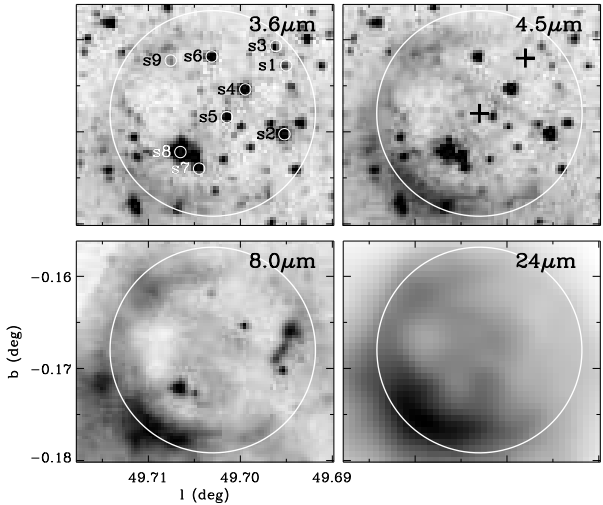}

\caption{Images of the central region at the indicated wavelengths.
Candidate ionizing stars are labeled from s1 to s9 on the 3.6 $\micron$
image. The large circles are centered on N103 and have a radius of
0.67$\arcmin$. Two crosses in 4.5 $\micron$ image are the center of N102
and N103.}

\label{shell_rgb_irac_cent}
\end{figure}

Figure \ref{shell_rgb_irac_mips} shows an enlarged view of this
double shell structure from the composite image of IRAC and MIPS
bands. The inside and outside shells are N103 and N102, respectively
\citep{Churchwell06}. A bright 8 $\micron$ shell (solid line in Figure
\ref{shell_rgb_irac_mips}$b$) encloses bright 24 $\micron$ emission
from the inner shell region. IRAC 8 $\micron$ mainly shows the emission
of polycyclic aromatic hydrocarbons (PAHs), which are excited in the
photodissociation region (PDR). Therefore PAHs emission trace the
ionization fronts well. MIPS 24 $\micron$ present the emission of the
small dust grains, which is inside of the shell. We note that these
shell structures are not visible in any of the 2MASS near-IR images,
despite their prominence in the GLIMPSE IRAC data. Average radii of
N102 and N103 are 2\farcm0 and 0\farcm67, corresponding to 3.3 pc and
1.1 pc, respectively.  Here we adopt a kinematic distance to the shells
as 5.7 kpc, which were deduced from the velocity of the associated
molecular material, $V_{LSR}=$ 59 \kmsc\ from our HHT observations,
and the Galactic rotation curve of \citet{Brand93}.

This shell structure lies in the direction of the major star-forming
regions in the Galaxy \citep{Avedisova02}. This apparent association
on the plane of the sky does not prove that the shell results from the
expansion of an \ion{H}{2} region caused by star formation activity,
rather than some other process (e.g., supernova remnant, planetary nebula,
etc.). To determine whether our double shell represent is \ion{H}{2}
region or not, we compile the radio continuum observations at various
frequencies:
$S_{\rm4.875\,GHz}$ = 0.55 Jy and $S_{\rm14.8\,GHz}$ = 0.21 Jy \citep{Wink82}, 
$S_{\rm1.4\,GHz}$ = 0.62 Jy \citep{Condon98}, 
$S_{\rm4.85\,GHz}$ = 0.33 Jy \citep{Gregory96}, 
$S_{\rm327\,MHz}$ = 0.80 Jy \citep{Taylor96}, 
$S_{\rm4.875\,GHz}$ = 0.55 Jy \citep{Altenhoff79}.

\begin{figure}
\epsscale{1.1}
\plotone{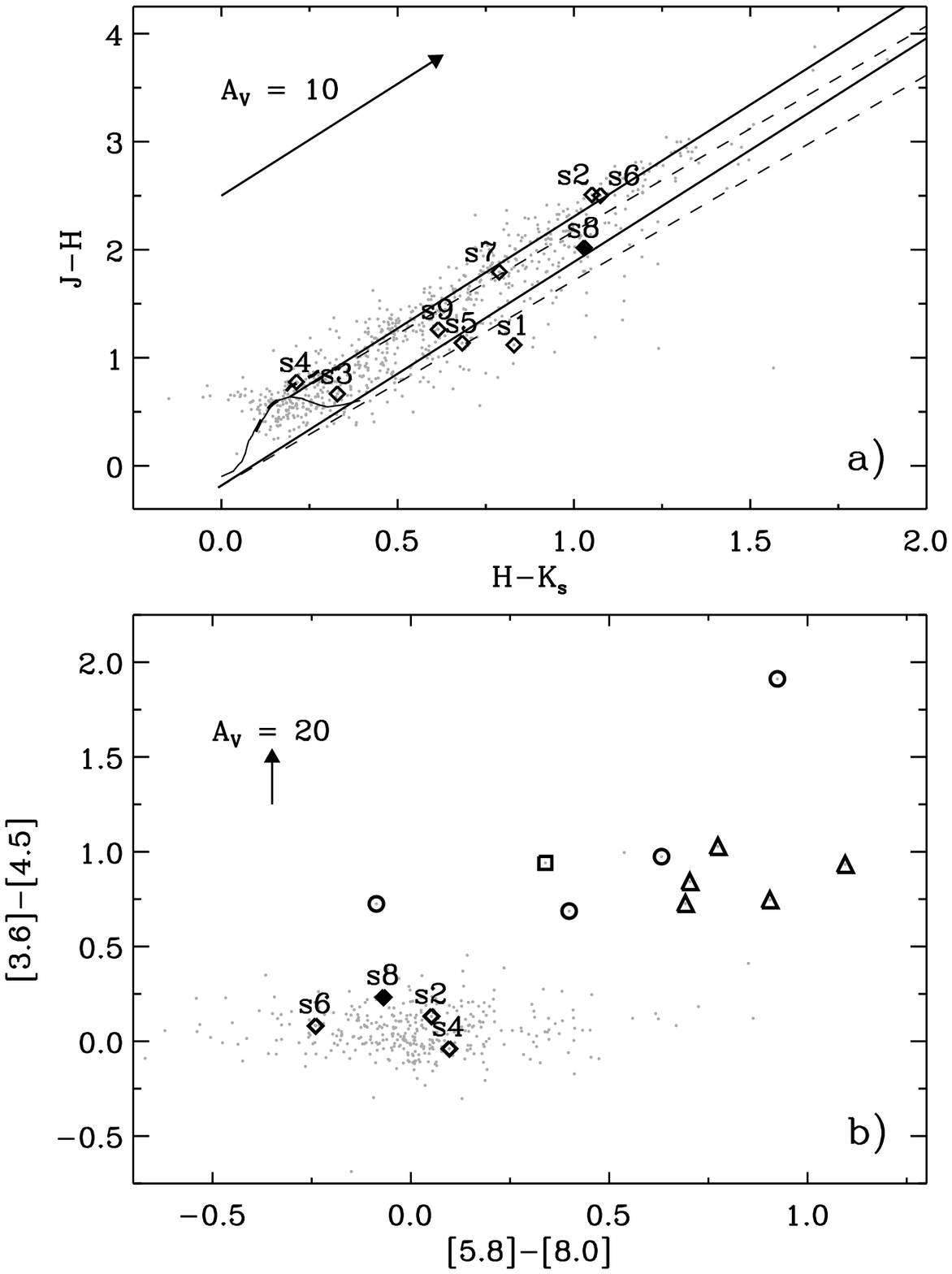}

\caption{ ({\it a}) JHKs color-color plot. The solid and dashed
curves represent the locus of main-sequence and giant stars
\citep{Bessell88}. The parallel solid and dashed lines, for O6-8V
\citep{Koornneef83} and M0V \citep{Bessell88}, are reddening vectors with
color excess ratios of 2.07 \citep{Kim07} and 1.9 \citep{Okumura00},
respectively. Reddening vector shows A$_V$=10 mag. ({\it b}) IRAC
$[3.6]-[4.5]$ vs. $[5.8]-[8.0]$ color-color plot. Gray dots are all
sources detected at each band through the line of sight in the shell
region. Open diamonds are stars within the radius of N103 seen in Figure
\ref{shell_rgb_irac_cent}. Filled diamond represents a candidate ionizing
star. YSO candidates are marked as circles for Stage 0/I, triangles
for Stage II, and squares for ambiguous sources. Reddening vector for
A$_{V}$=20 mag based on the extinction laws of \cite{Indebetouw05}
is shown as a filled arrow.}

\label{shell_c58_c34}
\end{figure}
\begin{deluxetable*}{rcrrrrrrrrrr}[!b]
\tabletypesize{\scriptsize}
\tablecaption{Model parameters for YSO candidates\tablenotemark{a}}
\tablewidth{0pt}
\tablehead{
\colhead{Number} & \colhead{Name} &
\colhead{$(\chi^2)$\tablenotemark{b}} &
\colhead{$L_{tot}$}       & \colhead{$\Delta
L_{tot}$\tablenotemark{c}} &
\colhead{$M_{\star}$}     & \colhead{$\Delta
M_{\star}$\tablenotemark{c}} &
\colhead{$\dot{M}_{env}$} & \colhead{$\Delta
\dot{M}_{env}$\tablenotemark{c}} &
\colhead{$A_{V}$}         & \colhead{$\Delta A_{V}$\tablenotemark{c}}
& {Stage}\\
&\colhead{(G$l+b$)}&&\colhead{($L_\odot$)}&\colhead{($L_\odot$)}&\colhead{($M_\odot$)}&\colhead{($M_\odot$)}&
   \colhead{($M_\odot yr^{-1}$)}&\colhead{($M_\odot
yr^{-1}$)}&\colhead{(mag)}&\colhead{(mag)}&}
\startdata

 1& G49.6510-0.1541&   0.01&      92&      99&   3.1&  1.0&
4.29$\times10^{-6}$&    2.97$\times10^{-5}$&  18.8&   9.1&    II\\
 2& G49.6548-0.1291&   1.30&    7408&    2937&  10.3&  1.3&
1.30$\times10^{-6}$&    4.83$\times10^{-5}$&  47.5&   3.1&    II\\
 3& G49.6554-0.1452&   0.00&      35&      37&   3.1&  0.8&
2.86$\times10^{-6}$&    1.27$\times10^{-5}$&   1.4&   1.8&    II\\
 4& G49.6608-0.1253&   2.10&     321&     625&   2.4&  2.4&
5.17$\times10^{-5}$&    1.39$\times10^{-4}$&  30.6&  21.3&   0/I\\
 5& G49.6608-0.1458&   0.39&    2143&    3248&   8.0&  3.8&
9.58$\times10^{-4}$&    1.28$\times10^{-3}$&   6.4&   7.1&   0/I\\
 6& G49.6657-0.1941&   0.11&     140&     236&   3.6&  1.1&
7.74$\times10^{-6}$&    2.85$\times10^{-5}$&  10.7&   4.6&   Amb\\
 7& G49.6698-0.2161&   0.26&      26&      24&   2.2&  1.2&
1.36$\times10^{-5}$&    2.74$\times10^{-5}$&   4.4&   2.3&   0/I\\
 8& G49.6713-0.1205&   0.17&      38&      31&   3.3&  0.4&
1.03$\times10^{-7}$&    2.91$\times10^{-7}$&   0.6&   0.8&    II\\
 9& G49.6776-0.1099&   2.59&     114&      52&   3.4&  0.4&
4.45$\times10^{-8}$&    1.83$\times10^{-6}$&  10.1&   2.7&    II\\
10& G49.7256-0.0873&  15.05&     262&      97&   4.2&  0.3&
1.58$\times10^{-8}$&    4.94$\times10^{-9}$&   3.0&   1.0&    II\\
11& G49.7261-0.0777&   0.98&      71&      59&   3.1&  0.8&
1.01$\times10^{-6}$&    7.95$\times10^{-6}$&  22.7&   9.0&    II\\
12& G49.7309-0.2173&   0.35&      28&      47&   1.9&  1.1&
7.29$\times10^{-6}$&    1.98$\times10^{-5}$&   1.9&   1.5&   Amb\\
13& G49.7343-0.1109&   0.07&    1218&    2673&   6.2&  3.2&
2.53$\times10^{-4}$&    5.70$\times10^{-4}$&  11.6&  10.3&   0/I\\
14& G49.7355-0.1695&   1.62&     337&     151&   5.9&  0.9&
8.28$\times10^{-5}$&    1.53$\times10^{-4}$&  19.1&   7.3&   0/I\\
15& G49.7409-0.1125&   0.01&     316&     136&   4.4&  0.6&
3.41$\times10^{-7}$&    1.12$\times10^{-5}$&  15.4&   2.8&    II\\
16& G49.7421-0.1766&   4.52&      52&      64&   2.6&  1.2&
1.59$\times10^{-5}$&    3.77$\times10^{-5}$&   6.9&   3.1&   Amb\\
17& G49.7470-0.2130&   0.71&      52&      49&   3.0&  1.0&
2.09$\times10^{-5}$&    5.31$\times10^{-5}$&   4.9&   1.9&   Amb\\
18& G49.7552-0.0775&   2.94&      95&      51&   3.6&  0.8&
6.72$\times10^{-5}$&    8.50$\times10^{-5}$&  16.8&  10.4&   Amb\\
19& G49.7637-0.0742&   0.00&      83&     176&   3.1&  0.9&
5.06$\times10^{-6}$&    1.60$\times10^{-4}$&  14.3&   9.2&    II

\enddata
\tablenotetext{a}{ The 0.152\ad $\times$ 0.150\ad region centered on
(l,b) =(49.696\ad, $-$0.148\ad)}
\tablenotetext{b}{ The $\chi^2$ value for the best-fitting SED. A good
fit($\chi^2/N_{data} \leq
4$) can have a $\chi^2$ value as high as 16 or more.}
\tablenotetext{c}{ The uncertainties on the luminosities, masses, mass
accretion rates, and extinctions
are calculated as the weighted standard deviation
of the  luminosities, masses, mass accretion rates, and extinctions of
all the acceptable YSO models.\\}
\end{deluxetable*}

\cite{Taylor96} concluded that most sources detected by the Westerbork
Synthesis Radio Telescope (WSRT) at 327 MHz and in the IRAS Point Source
Catalog are thermal radio sources associated with compact \ion{H}{2}
regions.  Their W1921+1442 is coincident with the shell structure as
seen in Figure \ref{shell_rgb_irac_mips} ({\it a}). IRAS fluxes of
the source are 1.0, 24.9, 225.8, and 292.5 Jy at 12, 25, 60, and 100
$\micron$. \cite{Taylor96} selected a ratio of $S_{\rm60\,\mu m}$ to $
S_{\rm 327\,MHz}$ = 200 as the boundary separating thermal and non-thermal
radio sources. The ratio of $S_{\rm60\,\mu m}$ to $ S_{\rm 327\,MHz}$
of W1921+1442 is greater than 200, consistent with thermal emission. The
IRAS flux ratio also lies in the region $0.5 < \log(S_{60}/S_{25})<1.7$
and $0.5 < \log(S_{25}/S_{12})<1.4$ occupied by compact \ion{H}{2} regions.

We use equation (1) of \cite{Simpson90} to estimate from the
radio-continuum flux the number of ionizing Lyman continuum photons
$N_{\rm LyC}$ emitted by the exciting star. We assume a fraction of
helium recombination photons to excited states $f_i = 0.65$, an electron
temperature $T_e = 10000$ K and a distance $D=5.7$ kpc. The integrated
flux density at 4.85 GHz is 0.33 Jy \citep{Gregory96}, indicating an
ionizing flux of log($N_{\rm LyC}) \sim 47.96$ photons per second. This
corresponds to a star of spectral type O9V \citep{Martins05}. The
value $S_{\rm4.875\, GHz}$ = 0.55 Jy \citep{Altenhoff79,Wink82} gives
log($N_{\rm LyC}) \sim 48.18$ photons per second, which implies on
O8.5V spectral type. The ionizing flux may be a lower limit if there is
significant dust absorption in the ionized gas.

\subsection{Identifying Ionizing Star}
\label{sec:ionizing_star}

Given that the N102 and N103 structures are shells swept up by expanding
\ion{H}{2} regions, we try to identify the ionizing stars using NIR
photometry. N102 (outer shell) shows a nearly complete circular shape
except for the missing region to the north-west which suggests an eruption
of the bubble. Ionizing star(s) might be located near the center of the
shell where the inner shell (N103) is located. Therefore, we consider
stars detected at all 2MASS bands within the inner shell as possible
candidates for ionizing star(s). Figure \ref{shell_rgb_irac_cent} shows
the candidate ionizing stars, labeled from s1 to s9, at different IRAC
and MIPS bands.

The most likely candidate to be the ionizing source was selected as
follows: Using photometry data from the 2MASS and Spitzer catalogs,
we placed sources in Figure \ref{shell_rgb_irac_cent} on color-color
diagrams (see Figure \ref{shell_c58_c34}). Extinction is clearly
severe toward the Sagittarius spiral arm tangency at $45\arcdeg <
l < 51\arcdeg$ including W51, since there is no detectable increase in the
star count \citep{Benjamin05,Churchwell09}. \cite{Okumura00} and
\cite{Kim07} estimated a color excess ratio $E_{J-H}/E_{H-K_{s}}$
of 1.9 and 2.07 toward W51A and W51B, respectively. In the $J-H$
vs. $H-K_s$ color-color diagram, we assumed a color excess ratio of
2.07. Although the color excess ratio is measured from the W51B region,
the value 2.07 traces the sources in the shell region better than 1.9
measured in W51A (see Figure \ref{shell_c58_c34}$a$). This excess ratio
is at the upper range of $1.73\pm0.40$ found in the Galactic reddening law
study of \cite{Indebetouw05}. Following the criterion by \cite{Kim07},
we rejected sources with $J-K_s < 2.2$ as foreground sources. Notice how
this includes source s4, which despite being located near the center
of the N102 structure, presents negligible extinction and has a color
consistent with the locus of giant stars by \cite{Bessell88}; moreover,
the source is easily detectable in optical plates from the Digitized
Sky Survey (DSS), which would exclude it from being a deeply embedded
source. Foreground sources s1 and s3 are also marginally detected in
the DSS optical band. Sources s2, s6 and s7, have $J-K_s > 2.2$ showing
significant reddening, but they are located in the region of the diagram
corresponding to the extension of the cool Giant locus towards highly
reddened values, and thus we also discard them as candidate ionizing
sources. The remaining source, s8, is located in a region of the diagram
likely to be occupied by deeply embedded O type stars.

Candidate ionizing stars detected in all IRAC bands are located
in the group of main-sequence and giant stars around (0,0) in the
Spitzer $[3.6]-[4.5]$ vs.\,$[5.8]-[8.0]$ color-color diagram in Figure
\ref{shell_c58_c34} ({\it b}), indicating that they are not YSOs.

\subsection{YSO Candidates}
\label{sec:YSO}

In order to determine whether the expanding \ion{H}{2} region is able
to trigger the star formation, we investigate if there are newly formed
stars in the vicinity of the shell structure. To find YSO candidates
associated with the shell structure we use the 2MASS and GLIMPSE
point source catalogs. There are 2189 sources in $0.152\arcdeg \times
0.150\arcdeg$ region around N102. We identity a total of 19 YSOs near
the shells using the SED fitter of \cite{Robitaille07} following the
procedures described in \citet{Povich09}. The SED fitter includes a grid
of 200,000 YSO model spectral energy distributions (SEDs) and finds a
model fit using a $\chi^2$-minimization. For SED fitting, we use the
sources detected in at least four bands ($N_{data} \geq 4$) of the 8 IR
bands, consisting of the 4 IRAC bands and the 3 2MASS bands for sources
with firm 2MASS identifications in the GLIMPSE Point Source Catalog,
and the MIPS 24 $\micron$ flux from the MIPSGAL survey if detected (see
Section \ref{sec:observation}). First, to remove stellar sources, we
fit stellar photosphere SEDs to all the sources with $N_{data} \geq 4$.
Sources with $\chi^{2}/N_{data} \leq 4$, indicating a good stellar fit,
are removed. Because the fitting tool take into accounts extinction
\citep{Indebetouw05}, even highly reddened stars should be removed by
this step. Asymptotic giant branch (AGB) stars are removed by fitting
AGB star SED templates included in the SED fitting tool.

\begin{figure}
\epsscale{1.2}
\plotone{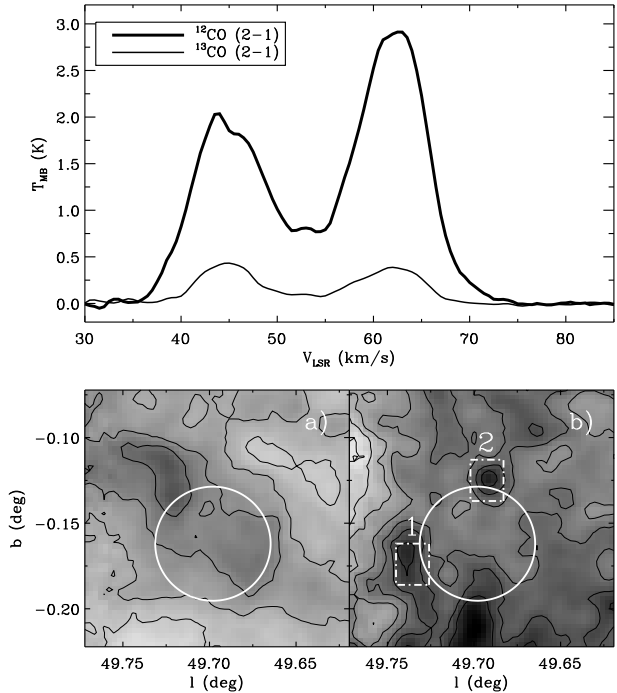}

\caption{{\it Top}: Average \co and \tco (J=2-1) spectra seen in
Figure
\ref{shell_rgb_irac_mips}. {\it Bottom} ({\it a}) The \co (J=2-1)
intensity integrated between 40 and 50 \kmsc. ({\it b}) The \tco
(J=2-1) intensity integrated between 56 and 67 \kmsc. The gray scale
is from 0 to 44 K \kmsc. Contour levels are 5, 10, 15, 20, 25, 30,
and 35 K \kmsc. Circles show the shell size of N102. The dash-dotted
rectangles labeled 1 and 2 show the fields of clumps 1 and 2 in Figure
\ref{shell_spectra_c1c2}.}

\label{shell_coint_4050_5667}
\end{figure}

Next, we fit the sources which are poorly fit by stellar photospheres
to the YSO models of \cite{Robitaille06}. We allow the distance range to
be from 5 to 9 kpc and the interstellar extinction from 0 to 60 mag in V
band for fitting YSOs. In the same manner as fitting stellar photospheres,
we consider sources with $\chi^{2}/N_{data} \leq 4$ to be good fits. We
discard sources not detected at 24 $\micron$ which show 8 $\micron$
flux excesses above an extrapolation of the 3 shorter wavelength IRAC
bands. These sources are usually contaminated at 8 $\micron$ by a noise
peak or a diffuse background feature.

\begin{figure}
\epsscale{1.1}
\includegraphics[height=0.348\textwidth]{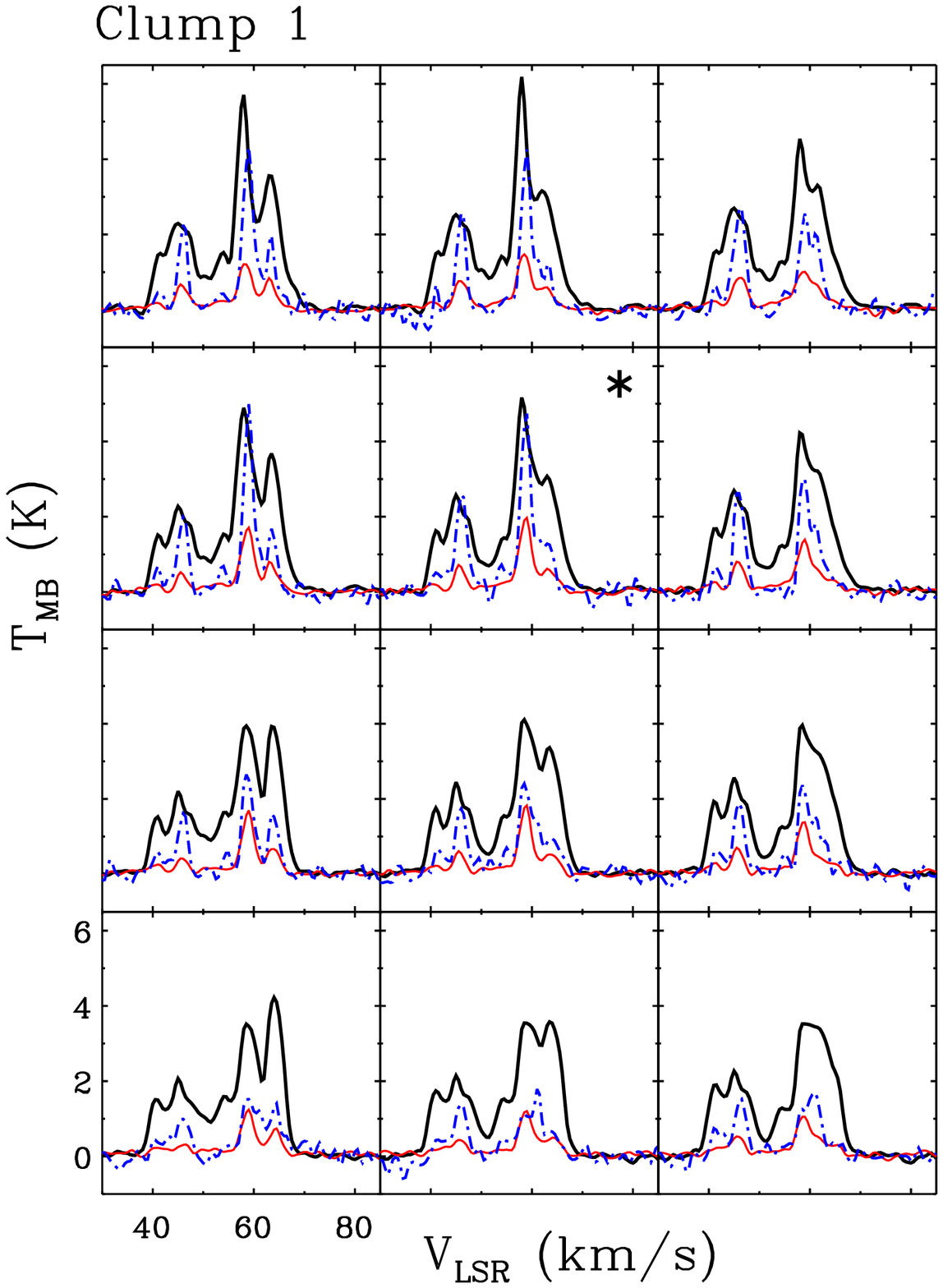}
\includegraphics[height=0.348\textwidth]{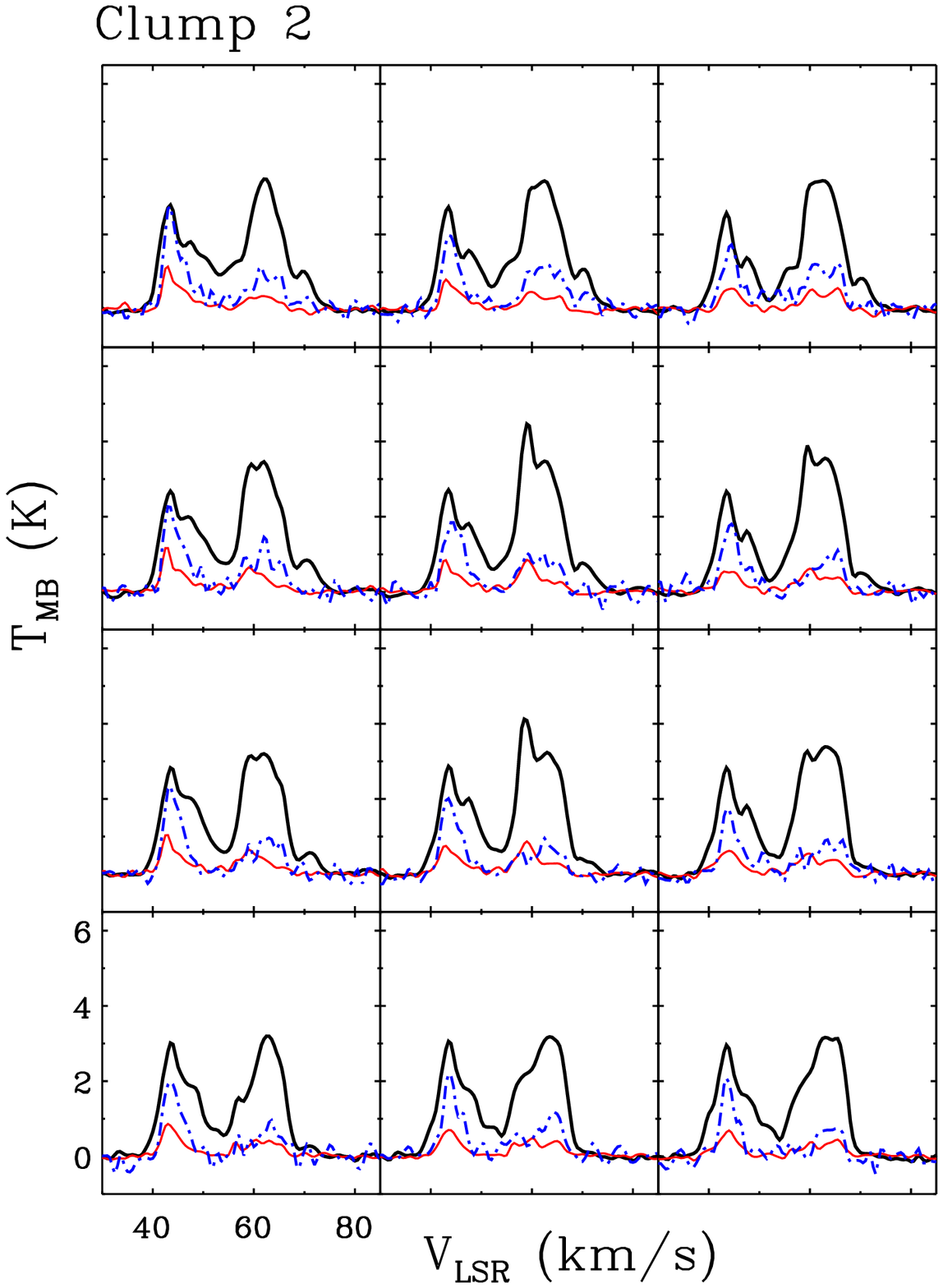}

\caption{\co (J=2-1) ({\it thick solid line}), \tco (J=2-1) ({\it
thin solid line}) and \tco (J=1-0) (dash-dotted line) spectra of clumps
1({\it left}) and 2 ({\it right}) seen in the dash-dotted boxes in Figure
\ref{shell_coint_4050_5667}. The spectra have 22$\arcsec$ spacing.}

\label{shell_spectra_c1c2}
\end{figure}

Each YSO stage is defined by a disk mass $M_{\rm disk}$ and an envelope
accretion rate $\dot{M}_{\rm env}$ provided by the SED fitting algorithm:
Sources are classified as Stage 0/I with $\dot{M}_{\rm env}/M_{\star} >
10^{-6}~{\rm yr}^{-1}$, Stage II YSOs with $\dot{M}_{\rm env}/M_{\star}
< 10^{-6}~{\rm yr}^{-1}$ and $M_{\rm disk}/M_{\star} > 10^{-6}$, and
Stage III with $\dot{M}_{\rm env}/M_{\star} < 10^{-6}~{\rm yr}^{-1}$ and
$M_{\rm disk}/M_{\star} < 10^{-6}$ \citep{Robitaille06}. We determine
the evolutionary stage of each source using the relative probability
distribution for the Stages of all the good-fit models. Note that we
use the term ``Stage'' to distinguish the evolutionary state of the
theoretical model from the term ``Class'' which derives from the observed
SED \citep{Evans09}. The good-fit models of each source are defined by
\begin{equation} 
	\chi^2 - \chi^2_{\rm min} \leq 2N_{\rm data}, 
\end{equation} 
where $\chi^2_{\rm min}$ is the goodness-of-fit parameter for the
best-fit model. The relative probability of each good-fit model is
estimated according to
\begin{equation}
	P(\chi^2)=e^{-(\chi^2-\chi^2_{\rm min})/2} 
\end{equation} 
and is normalized. After a probability distribution for the evolutionary
Stage of each source is constructed from the Stages of all the good-fit
models, the most probable Stage of each source is determined by requiring
$\Sigma P({\rm Stage}) \geq 0.67$. If this condition is not satisfied,
then the Stage of the source is considered as ``Ambiguous''. Model
parameters of all YSOs are listed in Table 1.

Figure \ref{shell_c58_c34}{\it b} shows IRAC $[3.6]-[4.5]$
vs.\,$[5.8]-[8.0]$ color-color diagram of sources within the shell
region. Note that we show only sources that are detected in all IRAC
bands. The gray dots represent foreground or background stars. Candidate
ionizing sources seen in Figure \ref{shell_rgb_irac_cent} are marked
with black diamonds. Filled diamond represents the most probable
ionizing star. YSOs with IR excess are marked as circles for Stage
0/I, triangles for Stage II, and squares for Ambiguous. In the IRAC
color-color diagram, YSO candidates are clearly separated from stellar
sources around (0, 0) where candidate ionizing stars are located. In
the \S\,\ref{sec:discussion}, we will discuss YSO candidates associated
with shell.

\section{Molecular clouds}
\label{sec:CO}
\subsection{Spatial Distribution and Kinematics}
\label{sec:distribution_kinematics}

\begin{figure*}
\epsscale{0.71}
\plotone{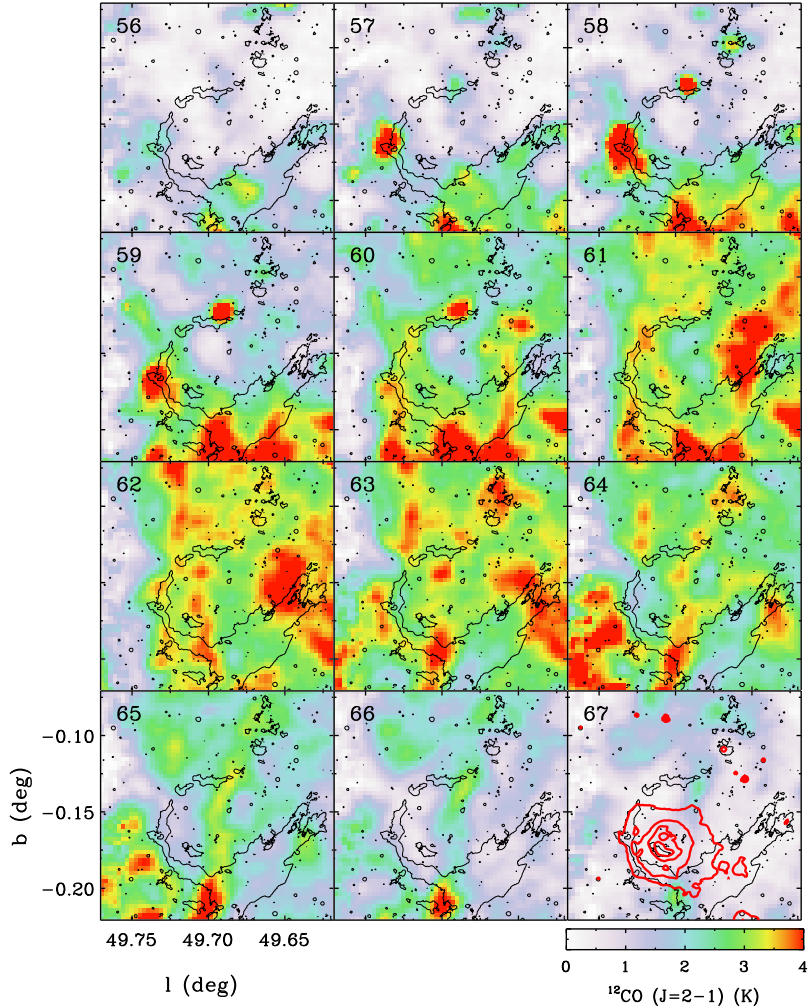}

\caption{The \co (J=2-1) channel maps by 1 \kms interval. Black
contour shows the shell structure of IRAC 8 $\micron$. Contour level
is 50 MJy sr$^{-1}$. In last channel map, red contours are MIPS 24
$\micron$. Contour levels are 50, 100, 200, and 400 MJy sr$^{-1}$. The
color scale bar is presented in units of K-T$_{\mathrm {MB}}$.}

\label{shell_co_chanmap}
\end{figure*}

To determine some physical properties of the shell structure, we used
the \co (J=2-1), \tco (J=2-1), and \tco (J=1-0) maps described in
Section 2. The first step is to determine the velocity range of the
molecular components associated with the shell structure. Figure
\ref{shell_coint_4050_5667} ({\it top}) shows the velocity
profile of each line averaged over the whole region shown in Figure
\ref{shell_rgb_irac_mips}. There are two bright components around 40
-- 50 \kms and 56 -- 67 \kms in the \tco(J=2-1) line. We present the
integrated intensity maps of \co (J=2-1) in two different velocity ranges
in bottom of Figure \ref{shell_coint_4050_5667}.

In the \co (J=2-1) map integrated from 40 to 50 \kms (Figure
\ref{shell_coint_4050_5667}{\it a}), a strong component is located on
the north-east side of the shell. Some bright emission is seen within
the radius of N102. The \co (J=2-1) intensity map integrated from
56 to 67 \kms (Figure \ref{shell_coint_4050_5667}{\it b}) shows that
some clumpy structures surround the ionized region seen in the IRAC
shell. We label two of these as clumps 1 and 2. The \co, \tco (J=2-1)
and \tco (J=1-0) spectra around clump 1 and 2 are presented in Figure
\ref{shell_spectra_c1c2}. Peak intensity of \co (J=2-1) appears around
59 \kms in both clumps. The \tco (J=2-1) line between 56 and 62 \kms
of clump 1 is very bright and distinguished from the other components
in different velocity ranges. The ratio, ${^{12/13}R_{2-1}}$, of \co
(J=2-1) and \tco (J=2-1) at the peak velocity of \tco (J=2-1) is 1.85
at the position marked as an asterisk in Figure \ref{shell_spectra_c1c2}
({\it left}). In all spectra of clump 1 and 2, the main beam temperature
of GRS \tco (J=1-0) is brighter than the main beam temperature of HHT \tco
(J=2-1) implying the non-LTE state of shell region (\S 4.2).

\begin{figure}
\epsscale{1.2}
\plotone{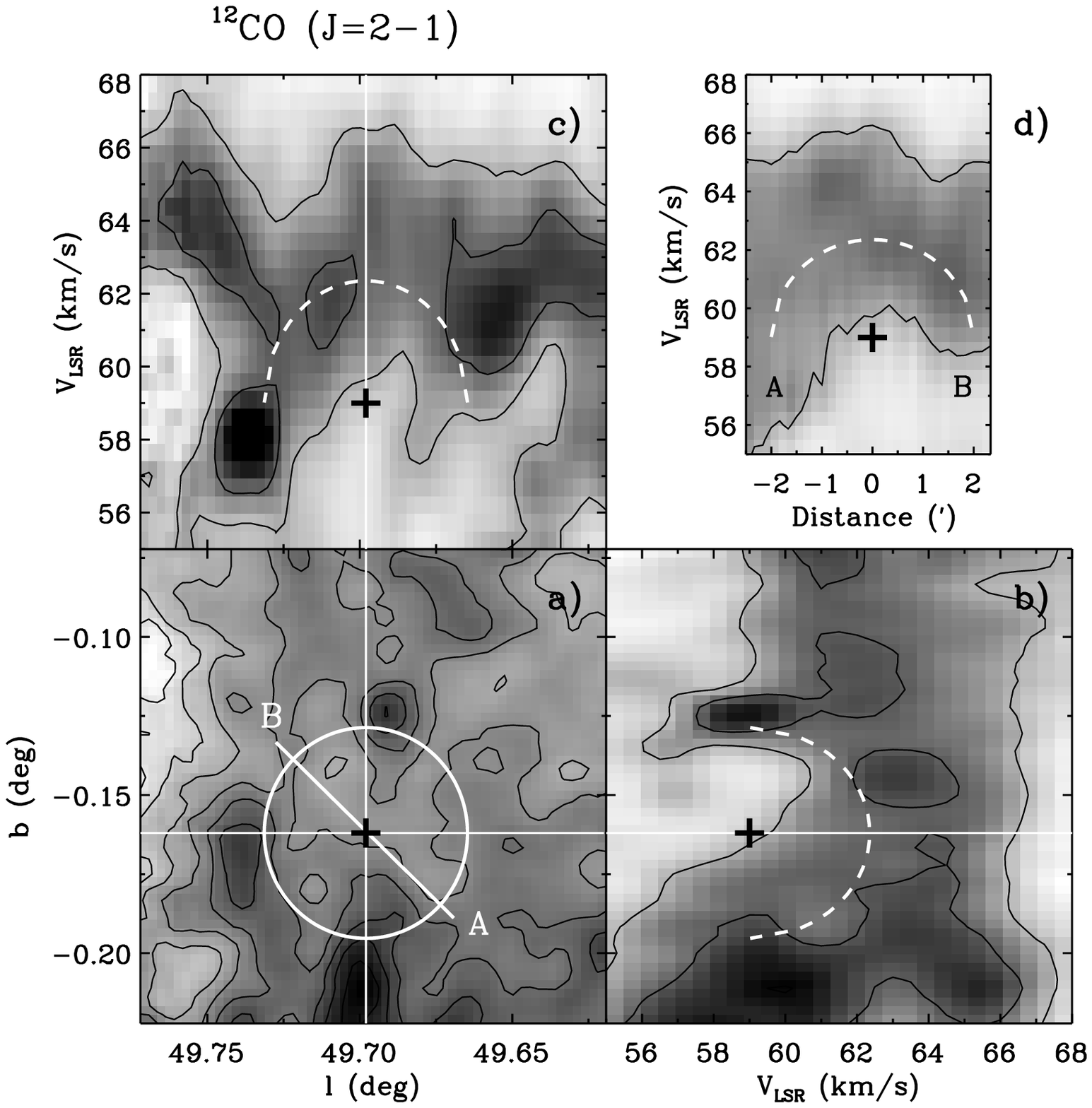}

\caption{\co spatial and position-velocity diagrams around the shell
region. ({\it a}) is the \co (J=2-1) intensity map integrated between 55
and 68 \kmsc. Contour levels are 5, 10, 15, 20, 25, 30, 35, 40 and 45 K
\kmsc. The cross symbol represents the center position of the N102 shell
listed by \cite{Churchwell06}. White circle represents the shell size of
2 arcmin radius. The oblique line AB shows the section for the PV map
presented in ({\it d}). ({\it b}) and ({\it c}) are position-velocity
maps in latitude and longitude across the solid lines in ({\it a}). The
contour levels are \{1, 2, 4\}$\times$1.6 K-T$_{\mathrm {MB}}~(5\sigma)$.
({\it d}) A position-velocity diagram along AB. The {\it{x}}-axis
is the distance from the center of the shell. Contour level is 1.6
K-T$_{\mathrm {MB}}$. The velocity variation expected from equation (1),
adopting $v_{0}$ = +3.4 \kmsc, $R$ = 2 arcmin, and $v_{sys}$ = +59 \kmsc,
is shown by the dashed line.}

\label{shell_pv}
\end{figure}

Figure \ref{shell_co_chanmap} shows \co (J=2-1) emission associated
with the shell between 56 and 67 \kms by 1 \kms steps. The molecular
and IRAC shells show good correspondence between 58 and 63 \kms in the
north-east quadrant where the shell is complete. To the north-west the
shell is broken in both the IRAC and molecular line images. To the south
the correspondence is less clear because of the presence of additional
molecular gas extending off the map in Figure \ref{shell_co_chanmap}. MIPS
24 $\micron$ emission represented in the last channel, 67 \kmsc, fills
the inside of the shell. The distribution of the CO molecular cloud
agrees with the broken shell structure seen in IRAC bands. A molecular
clump in the south forms the bright rim of the IRAC shell.

Figure \ref{shell_pv} shows the spatial and kinematic structure around
the shell in \co (J=2-1). Figure \ref{shell_pv}{\it a} is the \co (J=2-1)
intensity integrated between 55 and 68 \kmsc. Figures \ref{shell_pv}{\it
b} and \ref{shell_pv}{\it c} are position-velocity maps in latitude
and longitude across the solid lines in Figure \ref{shell_pv}{\it a},
respectively. Figure \ref{shell_pv}{\it d} is the position-velocity map
along a line $AB$ in the intensity map (Figure \ref{shell_pv}{\it a}).
The cross represents the center position of the N102 shell. Although
most CO emission is distributed broadly from 55 to 67 \kmsc, the shell
structure is visible in the velocity range $V_{LSR} \simeq$ 56 to
62 \kmsc.

For a simple shell structure, the expected velocity along the line of
sight can be expressed as
\begin{equation}
   v(X)=v_{0}\sqrt{1-\frac{X^2}{R^2}} + v_{sys}
\label{eqn:expansion}
\end{equation} 
where $v_{0}$ is the expansion velocity, $R$ is the radius of the shell
(2$\arcmin$) estimated from the IRAC images of N102, $X$ the projected
distance from the center of the shell, and $v_{sys}$ the systematic LSR
velocity of the whole system. We assume $v_{sys} = +59$ \kms since all
clumpy structures on the border of the shell show the peak at $v_{sys}
= +59$. The expansion velocity of the shell is determined to be $v_{0}
\simeq 3.4$ \kms by least-squares fitting for the mean velocity within the
shell radius. The expected velocity from equation (\ref{eqn:expansion})
is presented as a dashed line. Although molecular material around the
center of N102 has disappeared, especially toward north-west, molecular
gas far from the central star still remains. The dynamical time-scale of
N102 can be estimated by dividing the size, 2$\arcmin$, by the expansion
velocity $v_0$. Assuming a $v_{sys}$ between 58 and 59 \kms for N102, we
estimate an age of 0.71 to 0.96 Myr for the derived expansion velocity,
which varies from 3.4 \kms to 4.5 \kmsc.

\subsection{Physical Conditions}
\label{sec:physical_condition}

As mentioned in \S\ref{sec:distribution_kinematics}, the observed
ratio of \tco (J=2-1) and \tco (J=1-0), $^{13}R_{2-1/1-0}$$\sim$\,0.5, shows that the shell region is not dense enough to satisfy the
LTE condition. Considering the location of the shell region near the
active star-forming complex, W51A, the kinetic temperature must be
higher than 12K, derived from the maximum main beam temperature of \co
(J=2-1). We therefore need to use a non-LTE statistical equilibrium
treatment of the CO molecular excitation to study physical properties of
the molecular cloud. We use the escape probability radiative transfer
and photodissociation model of \cite{Kulesa05}, also described in
\cite{Povich09}, to calculate grids of CO level populations for wide ranges
of volume densities ($10^2 - 10^7$ cm$^{-3}$) and temperatures ($5 -
300$ K) assuming detailed balance and steady state. From these model
grids, the total CO column density at each observed pixel is computed
from the peak temperature, line widths and integrated intensities of
the observed CO lines. Assuming that the CO heating is dominated by
photon processes (e.g. the photoelectric heating of dust), a coarse
estimate of the incident radiation field can be made for each point in
the map. The photodissociation model is then applied to estimate a total
hydrogen column density from the CO data. This calculation is based on
the CO and H$_2$ photodissociation treatments of \cite{Dishoeck88} and
\cite{Black87}, respectively, using a total interstellar carbon abundance
of $C/H=2.4\times10^{-4}$ \citep{Cardelli96}. Because the CO abundance is
a strong function of column density, blind application of a uniform CO
``dark cloud'' abundance ($10^{-4}$) can lead to a gross underestimate of
the total hydrogen column and gas mass. The visual extinction $A_V$,
is calculated from $N(H)$ using $N(H)/A_V = 1.9 \times 10^{21}$ cm$^{-2}$
mag$^{-1}$ \citep{Bohlin78}.

To determine how much molecular material is swept by the expanding
\ion{H}{2} region, we calculate the column density of CO between
56 and 62 \kms centered on 59 \kms.  This range corresponds to the
velocities that the clumps 1 and 2 on the shell occupy on the PV map
(Figure \ref{shell_pv}).

\begin{figure}
\epsscale{1.2}
\plotone{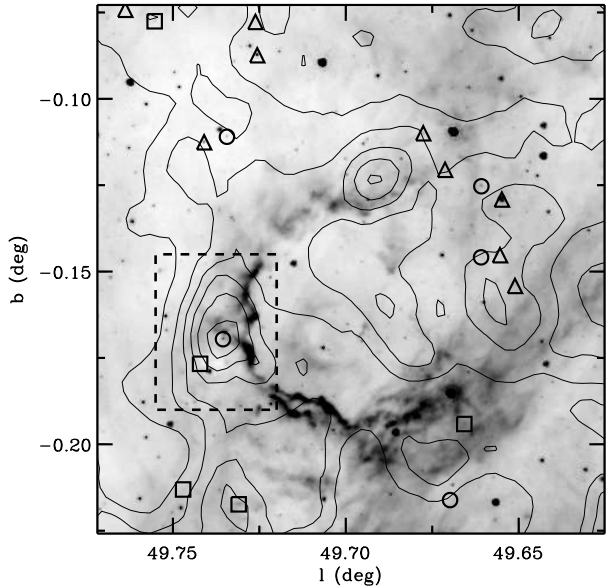}

\caption{Hydrogen column density contours superimposed on the IRAC
8.0 $\micron$ image (inverted grayscale). The hydrogen column density
is integrated over the velocity range between 56 and 62 \kmsc. Contour
levels are 40, 50, 60, 70, 80 and 90\% of the peak hydrogen column density
of $2.43 \times 10^{22}$ cm$^{-2}$. YSO candidates are marked as circles
for Stage 0/I, triangles for Stage II, and squares for ambiguous sources.
The dashed box shows the region in Figure \ref{shell_yso1416}.}

\label{shell_ir4_tcoint}
\end{figure}

We apply this analysis to the combination of \tco (J=1-0) data from GRS,
and the \co (J=2-1) and \tco (J=2-1) data from the HHT. We convolve all
spectral line maps to the GRS resolution of 46\arcsec. From the escape
probability model, we estimate a total mass of $3.2\times10^{4}~{\rm
M}_\odot$ distributed from 56 and 62 \kms in the $0.152 \arcdeg \times
0.150 \arcdeg$ region shown in Figure \ref{shell_rgb_irac_mips}.

To avoid degrading the resolution of our HHT maps, we also apply the
non-LTE model to the \co (J=2-1) and \tco (J=2-1) data. In this case,
the total mass in the same velocity range and region is computed to
be $2.9\times10^{4}~{\rm M}_\odot$. The derived total mass from two
lines of \co and \tco (J=2-1) is $\sim$ 10\% smaller than that derived
from three lines of \co and \tco (J=2-1) and \tco (J=1-0). Since two
results are similar, we present the total hydrogen column density $N(H)$
distribution derived from the HHT \co and \tco (J=2-1) with 32\arcsec\
spatial resolution in Figure \ref{shell_ir4_tcoint}.

\section{Discussion}
\label{sec:discussion}

Using the CO data and the identified YSO candidates, we investigate
whether our shell region are consistent with the predictions from the
star formation triggered by the expansion of an \ion{H}{2} region.
Figure \ref{shell_ir4_tcoint} shows the resulting distribution of the
total hydrogen column density $N(H)$ with the shell structure on the
IRAC 8 \micron\ image. YSO candidates are indicated by circles for Stage
0/I, triangles for Stage II, and squares for ambiguous sources. Total
hydrogen column density is derived from 56 to 62 \kmsc, which is the
velocity range affected by the expansion of the \ion{H}{2} region. Most
dense condensations are distributed along the 8 \micron\ shell. The
distribution of the hydrogen in the inner region of the large shell
is coincident with the distribution of the MIPS 24 \micron\ emission
encompassed by the IRAC 8 \micron\ shell. The distribution of gas
condensations along the 8 \micron\ shell, including clump 1 and 2
(see Figure \ref{shell_coint_4050_5667}), supports that the molecular
material collects on the boundary of the shell during the expansion of
the \ion{H}{2} region.

\begin{figure}
\epsscale{1.1}
\plotone{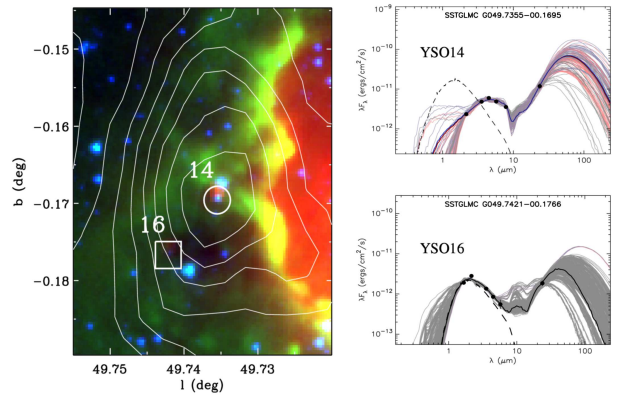}

\caption{$A_{V}$ contours superimposed on the composite image of MIPS
24 $\micron$ (red), IRAC 8.0 $\micron$ (green), and IRAC 4.5 $\micron$
(blue) of the dashed box region seen in Figure \ref{shell_ir4_tcoint}.
Contour levels are \{5, 6, 7, 8, 9, 10, 11, 12\} mag. The candidate YSOs
14 and 16 are marked with circle (Stage 0/I) and square (Ambiguous).
The SED model fits to the two sources are plotted on the right. The
dashed curve shows the reddened photosphere of the central star.}

\label{shell_yso1416}
\end{figure}

According to the ``collect and collapse'' model \citep{Elmegreen77},
a thin layer of compressed neutral material forms between the
ionization front and the shock front as the \ion{H}{2} region
expands. This layer may become gravitationally unstable, fragment and
form stars. \cite{Whitworth94} have shown that the shell collected by
the expansion of an \ion{H}{2} region will fragment when its column
density reaches $6 \times 10^{21}$ cm$^{-2}$.  Based on our
escape probability model using the \co and \tco (J=2-1) lines, the
hydrogen column density of the most shell region is already greater
than $1.0 \times 10^{22}$ cm$^{-2}$ (the lowest contour in Figure
\ref{shell_ir4_tcoint} ; 40\% of the peak hydrogen column density).
Note that this $N(H)$ are a lower limit because the CO molecule may be
depleted by freezeout onto dust grains.  Therefore by the criterion of
\cite{Whitworth94}, we conclude that the column density of the shell
region is sufficient to form a new star by the collect and collapse
process.

Dynamical expansion of the \ion{H}{2} region has been simulated by
\cite{Hosokawa06}. They describe the results of several models with
different central stars (12 - 101 M$_{\odot}$) and ambient densities
($10^{2}-10^{4}$ cm$^{-3}$). We estimate that the N102 region has an
O8.5V ionizing star with the ambient number density of about $10^{3}$
cm$^{-3}$, which we derive from $N(H)$ of $1.0 \times 10^{22}$
cm$^{-2}$ and a line-of-sight depth of 2 pc, corresponding to
the projected size of Clump 1.  We compare the physical properties of
N102 with the model S19 of \cite{Hosokawa06} : central star mass of 19
M$_{\odot}$ with an ambient density of $10^3$ cm$^{-3}$.  In this model,
a star produces a shell of radius 3.5 pc after 1 Myr and an
unstable region appears at about 0.5 Myr.  For the N102 shell, the current
radius of N102 is 3.3 pc at 5.7 kpc distance. Assuming a $V_{sys}$
of 59 \kms for N102, we estimate an expansion velocity of 3.4 \kms and
an expansion time of 0.96 Myr.  The shell size and the time scale are
in excellent agreement with those of model S19 by \cite{Hosokawa06}.
Therefore, we conclude that triggered star-formation on the border of
N102 is possible by the expansion of the \ion{H}{2} region.

Given that the accumulated molecular material is sufficient to form stars,
we find two YSOs (14 and 16) associated with the Clump 1 that is the
densest region along the shell structure.  The hydrogen column density
at the peak of clump 1 is $2.4\times 10^{22}$ cm$^{-2}$, corresponding
to the visual extinction of $\sim 14$ mag. The mass of clump 1 shown
in Figure \ref{shell_yso1416} is $2.3\times10^3$ M$_\odot$ based on the
escape probability modeling using \co and \tco (J=2-1) lines. This mass
are lower limits because the CO molecule may be significantly depleted
by freezeout onto dust grains in the core of the cloud.

Figure \ref{shell_yso1416} shows the $A_V$ distribution of clump 1 on
the MIPS 24 $\micron$ (red), IRAC 8.0 $\micron$ (green) and 4.5 $\micron$
(blue) composite images. YSO 14 and 16 are marked as a circle and square,
respectively. YSO 14 and 16, classified as Stage 0/I and Ambiguous,
lie close to the central region of clump 1. MIPS 24 $\micron$ emission
fluxes of 94 and 15 mJy are detected from YSO 14 and 16.  We also
show the SED model fits to these source in Figure \ref{shell_yso1416}.
YSO 14 is an intermediate YSO of $\sim 6$ M$_\odot$ with an accretion
rate of $\sim 8 \times 10^{-5} M_\odot~yr^{-1}$. YSO 16 has a mass of
$\sim 3$ M$_\odot$ with an accretion rate of $\sim 1.6 \times 10^{-5}
M_\odot~yr^{-1}$. While YSO 16 is classified as an ambiguous stage,
a early stage is indicated because the good-fit models include 62\%
(Stage 0/I), 20\% (Stage II) and 18\% (stage III).

If considering the total column density, age and size of the \ion{H}{2}
region predicted from the collect and collapse picture, star formation
should be possible everywhere the material is collected on the border
of the shell. As seen in Figure \ref{shell_ir4_tcoint}, although the CO
emission is collected on the periphery along the IRAC shell, we cannot
be sure whether all the YSOs seen in Figure \ref{shell_ir4_tcoint} are
associated with the material collected by the expansion of the \ion{H}{2}
region. At least, YSO 14 and 16 seem to be associated with the collected
material directly. We conclude that those two YSOs have formed probably
by the fragmentation of the collected material swept up by the expansion
of the \ion{H}{2} region.

It may be that the exciting star s8 which we have identified was itself
formed by such a triggering process. The molecular gas in the shell
has a very similar velocity to that of the gas associated with the W51A
\ion{H}{2} regions, so these clouds may be physically associated. If so,
we may speculate that expansion of the ionized regions in W51A triggered
the collapse leading to the formation of O-star s8 in a kind of sequential
star formation.

\section{Conclusions}
\label{sec:conclusion}

We have studied triggered star formation near the multiple shells of N102
and N103 \citep{Churchwell06} using Spitzer IR and HHT CO observations.
We have identified the candidate ionizing star (possibly O8.5V) of
the double shell. The CO observations confirm that dense molecular
material has been collected along the shell which has been detected in
the Spitzer IRAC images. The CO distribution shows that the shell is
blown out toward a lower density region to the northwest.  We find two
clumpy CO condensations with the same central velocity of 59 \kms to
the east and north along the edge of the IRAC shell.  Total hydrogen
column density around the shell is high enough to form new stars and we
identify two YSOs in early stages (Stage 0/I) within the densest molecular
clump 1.  Using the CO position-velocity map, we find that the \ion{H}{2}
region is currently expanding with a velocity of 3.4 \kmsc, suggesting
the shell's expansion time of $\sim$ 1 Myr.  This timescale is in good
agreement with numerical simulations of the expansion of the \ion{H}{2}
region \citep{Hosokawa06}.  We conclude that the star formation on the
border of N102 is triggered by the expansion of the \ion{H}{2} region.
\vspace{0.2cm}

\acknowledgments
We thank an anonymous referee for his/her thorough reading of the
manuscript and helpful comments.  We thank Matthew Povich and Marilyn
Meade for help with the SED fitter and MIPS photometry. M.\,K. also
thanks Amy Stutz, Kevin Flaherty, Seung-Hoon Cha and Yujin Yang for
helpful discussions. This research was supported in part by NSF grant
AST-0708131 to the University of Arizona. This work was supported by
the Korea Research Foundation Grant funded by the Korean Government
(MOEHRD: KRF-2007-612C00050).



\begin{thebibliography}{}
\bibitem[Altenhoff et al.(1979)]{Altenhoff79} Alternating, W.~J., Downes, D., Pauls, T., \& Schraml, J.\ 1979, \aaps, 35, 23
\bibitem[Avedisova(2002)]{Avedisova02} Avedisova, V.~S.\ 2002,Astronomy Reports, 46, 193
\bibitem[Bessell \& Brett(1988)]{Bessell88} Bessell, M.~S., \& Brett, J.~M.\ 1988, \pasp, 100, 1134
\bibitem[Benjamin et al.(2003)]{Benjamin03} Benjamin, R.~A., et al.\ 2003, \pasp, 115, 953
\bibitem[Benjamin et al.(2005)]{Benjamin05} Benjamin, R.~A., et al.\ 2005, \apjl, 630, L149 
\bibitem[Black \& van Dishoeck(1987)]{Black87} Black, J.~H., \& van Dishoeck, E.~F.\ 1987, \apj, 322, 412 
\bibitem[Bohlin et al.(1978)]{Bohlin78} Bohlin, R.~C., Savage, B.~D., \& Drake, J.~F.\ 1978, \apj, 224, 132
\bibitem[Brand \& Blitz(1993)]{Brand93} Brand, J., \& Blitz, L.\ 1993,\aap, 275, 67
\bibitem[Cardelli et al.(1996)]{Cardelli96} Cardelli, J.~A., Meyer, D.~M., Jura, M., \& Savage, B.~D.\ 1996, \apj, 467, 334 
\bibitem[Carey et al.(2005)]{Carey05} Carey, S.~J., et al.\ 2005, Bulletin of the American Astronomical Society, 37, 1252 
\bibitem[Churchwell et al.(2006)]{Churchwell06} Churchwell, E., et al.\ 2006, \apj, 649, 759
\bibitem[Churchwell et al.(2009)]{Churchwell09} Churchwell, E., et al.\ 2009, \pasp, 121, 213 
\bibitem[Condon et al.(1998)]{Condon98} Condon, J.~J., Cotton, W.~D., Greisen, E.~W., Yin, Q.~F., Perley, R.~A., Taylor, G.~B., \& Broderick, J.~J.\ 1998, \aj, 115, 1693 
\bibitem[Deharveng et al.(2003)]{Deharveng03} Deharveng, L., Lefloch, B., Zavagno, A., Caplan, J., Whitworth, A.~P., Nadeau, D., \& Mart{\'{\i}}n, S.\ 2003, \aap, 408,L25
\bibitem[Elmegreen \& Lada(1977)]{Elmegreen77} Elmegreen, B.~G., \& Lada, C.~J.\ 1977, \apj, 214, 725
\bibitem[Evans et al.(2009)]{Evans09} Evans, N., et al.\ 2009, arXiv:0901.1691 
\bibitem[Fazio et al.(2004)]{Fazio04} Fazio, G.~G., et al.\ 2004, \apjs, 154, 10 
\bibitem[Gregory et al.(1996)]{Gregory96} Gregory, P.~C., Scott, W.~K., Douglas, K., \& Condon, J.~J.\ 1996, \apjs, 103, 427
\bibitem[Hosokawa \& Inutsuka(2006)]{Hosokawa06} Hosokawa, T., \& Inutsuka, S.-i.\ 2006, \apj, 646, 240
\bibitem[Indebetouw et al.(2005)]{Indebetouw05} Indebetouw, R., et al.\ 2005, \apj, 619, 931
\bibitem[Kim et al.(2007)]{Kim07} Kim, H., Nakajima, Y., Sung, H., Moon, D.-S., \& Koo, B.-C.\ 2007, Journal of Korean Astronomical Society, 40, 17
\bibitem[Koo \& Moon(1997)]{Koo97} Koo, B.-C., \& Moon, D.-S.\ 1997, \apj, 475, 194
\bibitem[Koornneef(1983)]{Koornneef83} Koornneef, J.\ 1983, \aap, 128, 84
\bibitem[Kulesa et al.(2005)]{Kulesa05} Kulesa, C.~A., Hungerford, A.~L., Walker, C.~K., Zhang, X., \& Lane, A.~P.\ 2005, \apj, 625, 194
\bibitem[Kutner \& Ulich(1981)]{Kutner81} Kutner, M.~L., \& Ulich, B.~L.\ 1981, \apj, 250, 341
\bibitem[Martins et al.(2005)]{Martins05} Martins, F., Schaerer, D., \& Hillier, D.~J.\ 2005, \aap, 436, 1049
\bibitem[Okumura et al.(2000)]{Okumura00} Okumura, S.-i., Mori, A., Nishihara, E., Watanabe, E., \& Yamashita, T.\ 2000, \apj, 543, 799 
\bibitem[Pomar{\`e}s et al.(2009)]{Pomares09} Pomar{\`e}s, M., et al.\ 2009, \aap, 494, 987
\bibitem[Povich et al.(2009)]{Povich09} Povich, M.~S., et al.\ 2009, \apj, 696, 1278 
\bibitem[Rieke et al.(2004)]{Rieke04} Rieke, G.~H., et al.\ 2004, \apjs, 154, 25
\bibitem[Robitaille et al.(2006)]{Robitaille06} Robitaille, T.~P., Whitney, B.~A., Indebetouw, R., Wood, K., \& Denzmore, P.\ 2006, \apjs, 167, 256
\bibitem[Robitaille et al.(2007)]{Robitaille07} Robitaille, T.~P., Whitney, B.~A., Indebetouw, R., \& Wood, K.\ 2007, \apjs, 169, 328
\bibitem[Sault et al.(1995)]{Sault95} Sault, R.~J.,Teuben, P.~J., \& Wright, M.~C.~H.\ 1995, Astronomical Data Analysis Software and Systems IV, 77, 433
\bibitem[Simpson \& Rubin(1990)]{Simpson90} Simpson, J.~P., \& Rubin, R.~H.\ 1990, \apj, 354, 165
\bibitem[Skrutskie et al.(2006)]{Skrutskie06} Skrutskie, M.~F., et al.\ 2006, \aj, 131, 1163
\bibitem[Taylor et al.(1996)]{Taylor96} Taylor, A.~R., Goss, W.~M., Coleman, P.~H., van Leeuwen, J., \& Wallace, B.~J.\ 1996, \apjs, 107, 239
\bibitem[Watson et al.(2008)]{Watson08} Watson, C., et al.\ 2008, \apj, 681, 1341
\bibitem[Whitworth et al.(1994)]{Whitworth94} Whitworth, A.~P., Bhattal, A.~S., Chapman, S.~J., Disney, M.~J., \& Turner, J.~A.\ 1994, \mnras, 268, 291
\bibitem[Wink et al.(1982)]{Wink82} Wink, J.~E., Altenhoff, W.~J., \& Mezger, P.~G.\ 1982, \aap, 108, 227
\bibitem[van Dishoeck \& Black(1988)]{Dishoeck88} van Dishoeck, E.~F., \& Black, J.~H.\ 1988, \apj, 334, 771
\bibitem[Zavagno et al.(2006)]{Zavagno06} Zavagno, A., Deharveng, L., Comer{\'o}n, F., Brand, J., Massi, F., Caplan, J., \& Russeil, D.\ 2006, \aap, 446, 171
\end{thebibliography}
\end{document}